\documentclass[12pt]{article}
\usepackage[top=1.5cm, bottom=1.5cm, left=1.5cm, right=1.5cm]{geometry}
\usepackage{longtable}
\usepackage{natbib}
\usepackage{amsmath}
\usepackage{txfonts}
\usepackage{graphicx}
\usepackage{amssymb}
\usepackage[labelfont=bf]{caption}
\begin{document}
\title{Orbital period analysis of eclipsing Z Cam-type Dwarf Nova EM Cygni: Evidence of magnetic braking and a third body}
\author{Zhibin Dai $^{1,2}$, Shengbang Qian $^{1,2}$}

\footnotetext[1]{\scriptsize{National Astronomical Observatories/Yunnan
Observatory, Chinese Academy of Sciences, P. O. Box 110, 650011
Kunming, P. R. China.}}

\footnotetext[2]{\scriptsize{Key Laboratory for the Structure and Evolution of Celestial Objects, Chinese Academy of Sciences, P. R. China.}}

\maketitle

\begin{abstract}
\small

\noindent{Combining with our newest CCD times of light minimum of EM Cygni, all 45 available times of light minimum including 7 data with large scatters are compiled and the updated O-C analysis is made. The best-fit for the O-C diagram of EM Cygni is a quadratic-plus-sinusoidal fit. The secular orbital period decrease rate $-2.5(\pm0.3)\times10^{-11} s\;s^{-1}$ means that magnetic braking effect with a rate of mass loss via stellar wind, $2.3\times10^{-10} M_{\odot}yr^{-1}$, is needed for explaining the observed orbital period decrease. Moreover, for explaining the significant cyclical period change with a period of $\sim17.74(\pm0.01)yr$ shown in the O-C diagram, magnetic activity cycles and light travel-time effect are discussed in detail. The O-C diagram of EM Cygni cannot totally rule the possibility of multi-periodic modulation out due to the gaps presented after 25000 cycles. Based on the hypothesis of a K-type third star in literature, light trave-time effect may be a more plausible explanation. However, the low orbital inclination of the third body ($\sim7^{\circ}.4$) suggests that the hypothetic K-type third star may be captured by EM Cygni. But assuming the spectral contamination from a block of circumbinary material instead of a K-type third star, the third star may be a brown dwarf in case of the coplanar orbit with parent binary.}

\end{abstract}

\begin{bfseries}
\noindent{Stars: cataclysmic variables; Stars : binaries : eclipsing; Stars : individual (EM Cygni)}
\end{bfseries}


\section{Introduction}

EM Cygni is a typical dwarf novae and classified as the Z Cam type subclass, since the usually regular eruption cycle is occasionally interrupted by irregular fluctuations of brightness with a low-amplitude $\sim0^{m}.4$ \citep{dow93}. Although extensive photometric and spectroscopic studies have shown that EM Cygni is both an eclipsing binary and a double-lined spectroscopic binary \citep[e.g.][]{mum69,kra64}, its accurately absolute parameters are still a problem due to the contamination from a K2-5V third star along the line of sight, which is discovered by \cite{nor00}. \cite{rob74} and \cite{sto81} suggested its mass ratio is larger than unity. But \cite{nor00} pointed out that the previous works never considered the spectral contamination from a K-type third star, which leads to the radial velocity semi-amplitude for the mass donor star is increased from $135\;km s^{-1}$ to $202\;km s^{-1}$, and derived the mass ratio of EM Cygni is less than unity, which solved a problem that the mass transfer in EM Cygni can cause dynamically instability \citep{sto81}.

All orbital period analysis of EM Cygni have indicated that its orbital period probably is decreasing \citep{mum80,beu84,csi08}. But \cite{beu84} and \cite{csi08} suggested that its decrease rate was small and not significant. In the updated O-C diagram derived by \cite{csi08}, a secular orbital period decrease for EM Cygni cannot be significantly obtained due to the large scatters of the new 7 eclipse times. Since \cite{nor00} has discovered a K-type third star contributes approximately 16\% of the light from the system, a new O-C analysis for EM Cygni is important to verify the previous conclusion.

In this paper, 45 available times of light minimum from 1962 to 2008
are presented in Sect. 2, including a new data from our observation.
Sect. 3 deals with the O-C analysis for EM Cygni. Finally, the discussions of the possible mechanisms for orbital period change are made in Sect. 4.

\section{Observation of times of light minimum}

A new time of light minimum is obtained from our CCD photometric
observations with the PI VersArray 1300B CCD camera attached to the
2.4-m RC telescope at Gao Meigu (GMG) observational base of Yunnan
Observatory in China. It was carried out on November 25, 2008 in
white light at 10-s exposure time. Two nearby stars which have the similar brightness in the same viewing field of telescope are chosen as the comparison star and the check star, respectively. All images were reduced by using PHOT (measure magnitudes for a list of stars) of the aperture photometry package of IRAF. Since a small hump locates in the center of eclipse, which is similar to the light curves obtained by \cite{mum69}and \cite{csi08}, we removed this hump firstly and then used a parabolic fitting method to derive a new CCD time of light minimum. Based on the uncertainty of the parabolic fit, the error of our mid-eclipsing timing can be derived as $\sim0^{d}.000072$.

The large uncertainties in the mid-eclipse timings for the CVs with low inclination suggest that the times of light minimum of EM Cygni in literature obtained by using photoelectric method without given errors, should have lower accuracy. Considering that \cite{beu84} have derived the error of their photoelectric data as $0^{d}.0006$, we arbitrarily adopted the equal errors of $0^{d}.0006$ for the three photoelectric data observed in the similar observation conditions \citep{mum80}, and the larger errors of $0^{d}.001$ for the previous data. In addition, \cite{csi08} reported 8 new eclipse times including an amateur eclipse data in Heliocentric Julian Ephemeris Dates (HJED), which correspond to Ephemeris Time (ET), and all other data are in Heliocentric Julian Dates (HJD), which correspond to coordinated Universal time (UTC). However, for avoiding spurious trends in the orbital period analysis, we converted all 45 the eclipse timings in many different time systems to Barycentric Julian Dynamical Date (BJDD), which correspond to Terrestrial Dynamical Time (TDT), by using the code \citep{stu80}. Since the recent 7 eclipse times without given errors observed by \cite{csi08} present larger scatters ($\sim0^{d}.008$) than that of the previous data and the O-C diagram in their paper never show the completed scatters of the 7 data points, we supposed an uniform error of $0^{d}.003$ for the 7 eclipse times. All 45 available times of light minimum covering about half a century are listed in Table 1.

\section{Analysis of orbital period change}

For matching with the BJDD data we applied, the ephemeris in HJD derived by \cite{beu84} should be converted to BJDD,
\begin{equation}
T_{min}=BJDD\;2437882.8607(3)+0^{d}.290909155(25)E,
\end{equation}
which is used to calculate the O-C values of EM Cygni. After linear revision, the new epochs and average orbital period were derived as,
\begin{equation}
T_{min}=BJDD\;2437882.86141(18)+0^{d}.2909090099(33)E,
\end{equation}
with variance $\sigma_{1}=2^{d}.8\times10^{-3}$. Based on Eq 2, the
calculated O-C values are listed in column 6 of Table 1. Since the O-C diagram shown in Fig. 1 presents a possible cyclical period change and all the previous orbital period analyses implied that a secular change in orbital period may be presented in EM Cygni, we attempted to use a quadric-plus-sinusoidal ephemeris to fit the data. The weights of data points are calculated from the errors of the mid-eclipse times. The least-square solution leads to
\begin{eqnarray}
{\rm (O-C)_{1}}&=&-1^{d}.90(\pm0^{d}.34)\times10^{-3}+2^{d}.64(\pm0^{d}.29)\times10^{-7}E\nonumber\\
&&-3^{d}.69(\pm0^{d}.44)\times10^{-12}E^{2}+1^{d}.83(\pm0^{d}.28)\times10^{-3}\nonumber\\
&&\sin[0^{\circ}.01617(\pm0^{\circ}.00003)E+353^{\circ}.5(\pm8^{\circ}.8)],
\end{eqnarray}
with variance $\sigma_{2}=2^{d}.0\times10^{-3}$, which is almost half of $\sigma_{1}$. The resulting O-C diagram is shown in Fig. 1. The F-test used to assess the significance of quadratic and sinusoidal terms in Eq. 3 is suitable \citep{pri75}, as long as the form of parameter $\lambda$ is corrected to be
\begin{equation}
\lambda=\frac{(\sigma_{1}^{2}-\sigma_{2}^{2})/4}{\sigma_{2}^{2}/(n-6)},
\end{equation}
where $n$ is the number of data. Thus, a calculation gives
$F(4,39)=8.4$, which indicates that the fit is significant well above $99\%$ level. However, since the small quadric term in Eq. 3 implies that the quadric ingredient in the O-C diagram of EM Cygni may be not significant, a possible linear-plus-sinusoidal ephemeris for the O-C diagram of EM Cygni is estimated to be
\begin{eqnarray}
{\rm (O-C)_{1}}&=&-8^{d}.14(\pm2^{d}.78)\times10^{-4}+2^{d}.07(\pm0^{d}.44)\times10^{-8}E\nonumber\\
&&2^{d}.06(\pm0^{d}.26)\times10^{-3}\sin[0^{\circ}.018634(\pm0^{\circ}.000007)E+273^{\circ}.4(\pm8^{\circ}.6)],
\end{eqnarray}
with variance $\sigma_{3}=2^{d}.5\times10^{-3}$, which is larger than $\sigma_{2}$. In order to describe the significant level of the quadric term in Eq. 3, the parameter $\lambda$ of F-test describing by a similar formula as Eq. 4,
\begin{equation}
\lambda=\frac{(\sigma_{3}^{2}-\sigma_{2}^{2})}{\sigma_{2}^{2}/(n-6)},
\end{equation}
indicates $F(1,39)=18.3$. This means that the orbital period decrease in EM Cygni is significant well above 99\% level. In addition, the reduced $\chi^{2}$ value for Eq. 3 is calculated to be 1.29 and the time span of the whole data set is more than two complete periods. Thus, this cyclical period variations should be significant. After removing the quadric element in the top panel of Fig. 1, the cyclical period changes are shown clearly in the middle panel of Fig. 1. Although the F-tests and the reduced $\chi^2$ all suggest that a significant cyclical period change can be detected in the O-C diagram of EM Cygni, the gaps after 25000 cycles caused by the sparse data may mask a non-sinusoidal modulation. Therefore, the cyclical period change presented in the O-C diagram of EM Cygni may be a quasi-period change with a period of $\sim17.74(\pm0.01)yr$. According to Eq. 3, the orbital period decrease rate is calculated to be $-2.5(\pm0.3)\times10^{-11} s\;s^{-1}$, which is nearly two orders of magnitude larger than the previous works \citep[e.g.][]{mum80,csi08}.

\section{Discussion}

\begin{figure}
\centering
\includegraphics[width=9.0cm]{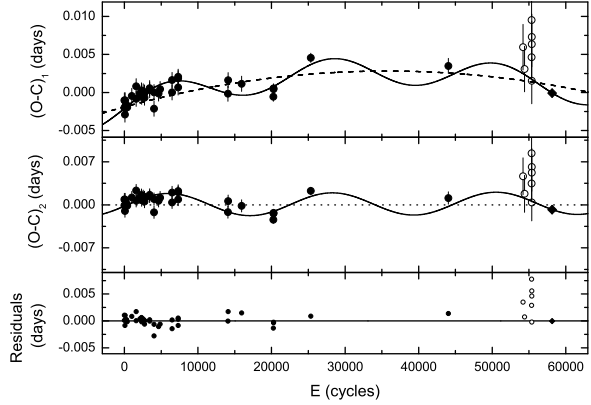}
\caption{The $(O-C)_{1}$ values of EM Cygni are fitted with the elements displayed in Eq. 3. The quadric and sinusoid curves are plotted in the dash and solid lines, respectively, in the top panel. After removing the quadric element, the $(O-C)_{2}$ is plotted in the middle panel. The solid circles and diamonds denote the data in literature and in our observation, respectively. The 7 eclipse times are plotted by open circles \citep{csi08}. The residuals and their linear fitted solid line are presented in the bottom panel.} \label{Fig. 1}
\end{figure}

\begin{figure}
\centering
\includegraphics[width=9.0cm]{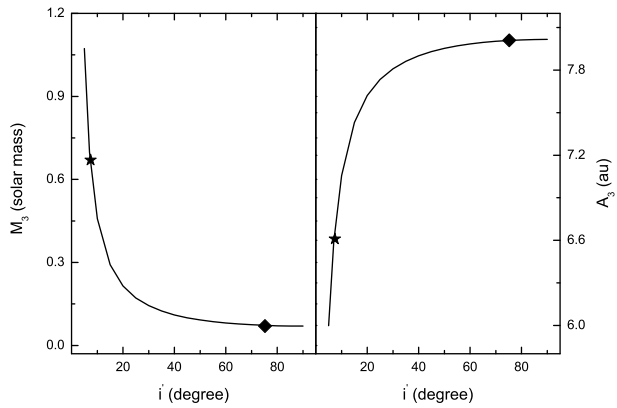}
\caption{The mass and separation of the third-body in EM Cygni depending on its orbital inclination $i^{'}$ are plotted in the left and right panels, respectively. The solid star and diamond in both panels refer to a K-type star and a brown dwarf, respectively.}
\label{Fig. 3}
\end{figure}

\subsection{Secular orbital period decrease}

The significant orbital period decrease shown in Fig. 1 implies a secular variation of orbital angular moment in EM Cygni. The updated spectroscopic studies indicated that the mass ratio of EM Cygni is less than unity \citep{nor00,wel07}. This means that the orbital period increase in EM Cygni is expected due to the mass transfer from the mass donor star to the white dwarf. Therefore, an available mechanism causing the observed orbital period decrease in EM Cygni should be considered.

In order to explain the observed period decrease, the decrease rate of orbital period of EM Cygni can be described as
\begin{equation}
\dot{P}=\dot{P}_{ml}+\dot{P}_{gr}+\dot{P}_{mt}+\dot{P}_{mb},
\end{equation}
where $\dot{P}_{ml}$, $\dot{P}_{gr}$, $\dot{P}_{mt}$ and $\dot{P}_{mb}$ are the orbital period change rates deduced by mass loss, gravitation radiation, mass transfer and magnetic braking, respectively. Since the orbital period of EM Cygni is long enough ($P\sim6.9h$), the gravitation radiation can be neglected \citep{ver81}. If we only discussed the conservative case (i.e. $\dot{P}_{ml}=0$), then the Eq. 7 can be described as
\begin{equation}
\dot{P}=\dot{P}_{mt}+\dot{P}_{mb}.
\end{equation}
Although the low orbital inclination of EM Cygni implies that the bright spot at the outer edge of the disk caused by the mass transfer may be invisible in eclipsing light curves \citep{mum69,mum80,sto81}, EM Cygni classified as a Z Cam subtype CV should have a high mass transfer rate, which is almost equal to the critical mass transfer rate \citep{war95}. By using the experiential formula \citep{sha86},
\begin{eqnarray}
{\rm \dot{M}_{crit}(M_{1},P)}&\sim&2.8\times10^{18}(1+0.11P/M_{1})^{3.47}\nonumber\\
&&\times[0.5-0.227log(0.11P/M_{1})]^{10.4}P^{1.73} gs^{-1},
\end{eqnarray}
where P and $M_{1}$ are the orbital period and mass of white dwarf, respectively, and the updated system parameters of EM Cygni \citep{wel07}, the critical mass transfer rate $\dot{M}_{crit}$ for EM Cygni is calculated to be $\sim1.2\times10^{-8} M_{\odot}yr^{-1}$. In addition, the accretion rate $5\times10^{-11} M_{\odot}yr^{-1}$ from the disk to the white dwarf derived by \cite{win03} can be neglected because it is nearly three orders of magnitude lower than the estimated $\dot{M}_{crit}$ of EM Cygni. Thus the mass transfer rate $\dot{M}_{mt}\sim\dot{M}_{crit}$ can produce the orbital period change with a rate of $\dot{P}_{mt}=8.6\times10^{-12} s\;s^{-1}$. A calculation with Eq. 8 leads to $\dot{P}_{mb}=-3.3\times10^{-11} s\;s^{-1}$. According to \cite{tou91}, the magnetic braking effect can be calculated by using
\begin{equation}
\frac{\dot{P_{mb}}}{P}=2(\frac{R_{A}}{a})^{2}\frac{M_{1}+M_{2}}{M_{1}M_{2}}\dot{M_{mb}},
\end{equation}
where $R_{A}$, a and $\dot{M}_{mb}$ are the Alfv\'{e}n radius of secondary, the separation of the binary and the stellar wind mass loss rate, respectively. Combining the orbital inclination $i=67^{\circ}$ and the orbital elements $a sin{i}=1.52\times10^{11}cm$ for EM Cygni \citep{nor00}, we estimated the separation of binary is about $2.37R_{\odot}$. Then $R_{A}^{2}\dot{M}_{mb}$ is derived as $5.1\times10^{-8} R_{\odot}^{2}M_{\odot}yr^{-1}$. Assumption that the Alfv\'{e}n radius of the K-type secondary is the same as that of the Sun(i.e. $\sim15R_{\odot}$), the mass loss rate caused by the stellar wind is estimated to be $2.3\times10^{-10} M_{\odot}yr^{-1}$.

\subsection{Quasi-period change}

\cite{hal89} has stated that the binaries with secondaries of later than F5 have cyclical period changes because of a solar-type magnetic activity cycle in the convective shell. As for EM Cygni, a K3-type secondary implies that a solar-type magnetic activity cycle in its convective shell for explaining the observed oscillations in the O-C diagram should be considered \citep{app92,lan06}. Although a calculation for EM Cygni has suggested that Applegate's mechanism cannot interpret the observed cyclical period variation, the fact that Applegate's model can be ruled out as a viable explanation does not mean that the hypothesis of period modulations driven by magnetic activity cycles in the secondary has to be discarded \citep{lan02,lan05,lan06}. Based on the gaps shown in Fig. 1, a magnetic activity cycles may be a well explanation for the quasi-period change in EM Cygni. However, \cite{lan06} also pointed out that the new improved model they proposed may be faced with a energy balance problem in the case of the secondary components of CVs due to their fast rotations. Moreover, the updated plot of mass ratio vs. secondary component's spectral type as the diagram in \cite{hal89} for almost all type of binaries has definitely rules the magnetic activity out as unique cause for a cyclical period variation \citep{lia10}. Considering the discovery of the spectrum from a K-type third star \citep{nor00}, it is more plausible that the light travel-time effect caused by a perturbations from a tertiary component is used to explain the observed cyclical variation. Since \cite{nor00} has detected the spectrum of a K-type third star in EM Cygni, the hypothesis of a strictly sinusoidal period change in its O-C diagram is expected. The sinusoidal fit presented in Fig. 1 suggests that the orbital eccentricity of the third star should be zero, and the amplitude of sinusoidal curve determines the projected distance $0.32(\pm0.05)AU$ from the third star to the mass center of the triple system. Then, according to the Third Kepler Law, the mass function of the third component $f(m_{3},i^{'})$ can be calculated to be $1.0(\pm0.5)\times10^{-6}M_{\odot}$. Upon that, it is worthy discussing whether the third body resulting in the cyclical period changes is the K2-5V third star derived by \cite{nor00} or not. In the following calculations, a combined mass of $1.0(\pm0.09)M_{\odot}+0.77(\pm0.08)M_{\odot}$ for the eclipsing pair of EM Cygni is used \citep{wel07}.

\subsubsection{a K-type third star}

Based on the calibration of MK spectral types, the mass of a K2-5V star can be estimated to be about $0.67M_{\odot}$, which implies that the orbital inclination of the third body is very low ($i^{'}\sim7^{\circ}.4$). And then the distance from the mass center of system is derived as $\sim6.6 AU$. In this case, the triple system is odd and the third body may be originated from a straggler, which is captured by EM Cygni in a long time ago and now has the similar systemic velocity to the parent binary \citep{nor00}. Moreover, this geometric structure of the triple system supports the prediction that the third star and the CV is at similar distances \citep{nor00}. Assuming the distances of EM Cygni is $\sim320pc$ \citep{beu84}, the separation from the third star to the parent binary $\sim9.1 AU$ projected on the sky can be estimated as $\sim0.03^{"}$, which is far beyond the resolution of the most advanced ground-based telescope. Therefore, this third star is hard to be separated from the parent binary. If this is the truth, then EM Cygni may be an important object for studying the dynamics of capturing a third body, and it is very interested to focus on the future evolution state of the K-type third star and the CV.

\subsubsection{a brown dwarf as the third body}

Since a K-type third star in EM Cygni suggests an unusual triple system, we can discuss another possible case by assuming the spectral contamination for EM Cygni is not from a K-type third star but from a block of circumbinary material along the line of sight. Thus, the observed cyclical period changes should be caused by a new third body, which may be a brown dwarf with a distance $\sim8.0 AU$ from the mass center of system shown in the right panel of Fig. 3, as long as $i^{'}>76^{\circ}$ (with a low possibility $<15\%$). This implies that a brown dwarf as the third body is almost coplanar with EM Cygni and can survive the previous evolution of the common envelope evolution of the parent binary. In this case, EM Cygni hiding a brown dwarf is an important system for further study of the formation and evolution of substellar objects as the dwarf nova Z Cha \citep{dai09}.

\section{Conclusion}

A new O-C diagram of EM Cygni by adding a new eclipse time with high precision has been made and detailed orbital period analysis implies that the best-fit of the O-C diagram shown in Fig. 1 is a sinusoidal variation with a period of $\sim17.74(\pm0.01)yr$ superimposed on the secular orbital period decrease with a rate of $2.5(\pm0.3)\times10^{-11} s\;s^{-1}$. Using the updated absolute parameters of EM Cygni \citep{wel07}, we proposed that magnetic braking effect may be a possible explanation for the observed secular orbital period decrease. According to the experiential formula \citep{sha86}, a critical mass transfer rate of Z Cam-type dwarf nova EM Cygni is estimated to be $\sim1.2\times10^{-8} M_{\odot}yr^{-1}$. Moreover, we first detected a cyclical period change in the O-C diagram of EM Cygni. Both plausible mechanisms that magnetic activity cycles and light travel-time effect are discussed. Although the gaps shown in Fig. 1 indicate that the former mechanism cannot be totally ruled out, the spectrum of a K-type third star detected by \cite{nor00} suggests that the latter one also a possible mechanism. Based on the parameters of the K-type third star, we derived an odd structure of the triple system for EM Cygni with an extremely low orbital inclination of the K-type third star ($i^{'}\sim7^{\circ}.4$), which implies that the K-type third star may have a different origin (is captured by EM Cygni in a certain evolution stage) with parent binary. However, if we assumed that the spectral contamination is not from a K-type third star, but instead from a block of circumbinary material, then the observed cyclical period change suggests that the third body may be a brown dwarf whilst it is nearly coplanar with EM Cygni. We noted that the derived error of our mid-eclipsing timing seems to be underestimated for EM Cygni. This means that the conclusions we have made are just temporary and the more observations are expected. Anyway, EM Cygni is a controversial object and deserves to further observing in a longer base line including photometries and spectroscopies with high precision.

\section*{acknowledgements}

\small{This work was partly Supported by Special Foundation of President of
The Chinese Academy of Sciences and West Light Foundation of The
Chinese Academy of Sciences, Yunnan Natural Science Foundation
(2008CD157), and Yunnan Natural Science Foundation (No. 2005A0059M)
and Chinese Natural Science (No.10573032, No. 10573013 and
No.10433030). CCD photometric observations of EM Cygni
were obtained with the 2.4-m telescope at Yunnan Observatory.We thank the referee very much for the helpful comments and suggestions that helped to improve this paper greatly.}


\begin{center}
\begin{longtable}{p{3cm}cccccc}
\caption{The 45 times of light minimum for the Z Cam-type dwarf nova EM Cygni.}\\
\hline\hline
\hspace{2em}BJDD & type & $error^{d}$ & Method & E (cycle) & $(O-C)^{d}$ & Ref.\\
\hspace{1.5em}2400000+&&&&&&\\
\hline
\endfirsthead
\caption{Continued.}\\
\hline\hline
\hspace{2em}BJDD & type & $error^{d}$ & Method & E (cycle) & $(O-C)^{d}$ & Ref.\\
\hspace{1.5em}2400000+&&&&&&\\
\hline
\endhead
\hline
\endfoot
\hline
\multicolumn{7}{p{12cm}}{\scriptsize{Note. $^{+}$ the given uncertainty in literature. * the unknown method used by amateur astronomer. pe and ccd means the photoelectric and CCD methods used in observations, respectively. References: (1) \cite{mum69}; (2) \cite{mum80}; (3) \cite{beu84}; (4) \cite{csi08}; (5) This paper.}}
\endlastfoot
37882.860400 & pri & .001    & pe  & 0     & -.001040  & (1)\\
37883.732200 & pri & .001    & pe  & 3     & -.002000  & (1)\\
37906.713100 & pri & .001    & pe  & 82    & -.002890  & (1)\\
37911.660400 & pri & .001    & pe  & 99    & -.001020  & (1)\\
37936.677900 & pri & .001    & pe  & 185   & -.001680  & (1)\\
37966.641400 & pri & .001    & pe  & 288   & -.001860  & (1)\\
37968.677900 & pri & .001    & pe  & 295   & -.001660  & (1)\\
37996.604900 & pri & .001    & pe  & 391   & -.001950  & (1)\\
38174.933600 & pri & .001    & pe  & 1004  & -.000470  & (1)\\
38345.698500 & pri & .001    & pe  & 1591  &  .000826  & (1)\\
38348.605900 & pri & .001    & pe  & 1601  & -.000843  & (1)\\
38496.970200 & pri & .001    & pe  & 2111  & -.000129  & (1)\\
38561.552400 & pri & .001    & pe  & 2333  &  .000273  & (1)\\
38562.424300 & pri & .001    & pe  & 2336  & -.000546  & (1)\\
38624.388600 & pri & .001    & pe  & 2549  &  .000134  & (1)\\
38674.715700 & pri & .001    & pe  & 2722  & -.000020  & (1)\\
38675.588400 & pri & .001    & pe  & 2725  & -.000050  & (1)\\
38676.751400 & pri & .001    & pe  & 2729  & -.000689  & (1)\\
38878.934400 & pri & .001    & pe  & 3424  &  .000561  & (1)\\
38883.879600 & pri & .001    & pe  & 3441  &  .000314  & (1)\\
39052.604400 & pri & .001    & pe  & 4021  & -.002120  & (1)\\
39054.642900 & pri & .001    & pe  & 4028  &  .000029  & (1)\\
39230.933600 & pri & .001    & pe  & 4634  & -.000122  & (1)\\
39293.770500 & pri & .001    & pe  & 4850  &  .000438  & (1)\\
39767.662600 & pri & .001    & pe  & 6479  &  .001630  & (1)\\
39769.697300 & pri & .001    & pe  & 6486  &  .000016  & (1)\\
40006.788800 & pri & .001    & pe  & 7301  &  .000674  & (1)\\
40007.953600 & pri & .001    & pe  & 7305  &  .001850  & (1)\\
40008.826500 & pri & .001    & pe  & 7308  &  .002050  & (1)\\
41980.605600 & pri & .001    & pe  & 14086 & -.000162  & (2)\\
41982.643700 & pri & .001    & pe  & 14093 &  .001620  & (2)\\
42515.879400 & pri & .001    & pe  & 15926 &  .001130  & (2)\\
43776.677400 & pri & .0006   & pe  & 20260 & -.000566  & (2)\\
43778.714900 & pri & .0006   & pe  & 20267 &  .000535  & (2)\\
43780.751100 & pri & .0006   & pe  & 20274 &  .000436  & (2)\\
45257.409400 & pri & .0006$^{+}$   & pe  & 25350 &  .004570  & (3)\\
50692.461300 & pri & .001    & pe  & 44033 &  .003480  & (4)\\
53650.426600 & pri & .003    &  *  & 54201 &  .005960  & (4)\\
53709.187300 & pri & .003    & ccd & 54403 &  .003090  & (4)\\
53984.393700 & pri & .003    & ccd & 55349 &  .009520  & (4)\\
53989.334200 & pri & .003    & ccd & 55366 &  .004630  & (4)\\
53990.500600 & pri & .003    & ccd & 55370 &  .007310  & (4)\\
53991.367500 & pri & .003    & ccd & 55373 &  .001530  & (4)\\
53993.408700 & pri & .003    & ccd & 55380 &  .006370  & (4)\\
54796.020231 & pri & .000072 & ccd & 58139 & -.000072  & (5)\\
\end{longtable}
\end{center}


\begin{thebibliography}{plainnat}

\bibitem[Applegate (1992)]{app92}
  Applegate, J. H., 1992, ApJ, 385, 621
\bibitem[Beuermann \& Pakull (1984)]{beu84}
  Beuermann, K., \& Pakull, M. W., 1984, A\&A, 136, 250
\bibitem[Csizmadia et al. (2008)]{csi08}
  Csizmadia, Sz., Nagy, Zs., \& Borkovits, T., et al., 2008, AN, 329, 39
\bibitem[Dai \& Qian (2009)]{dai09}
  Dai, Z. B., \& Qian, S. B., 2009, ApJ, 703, 109
\bibitem[Downes \& Shara (1993)]{dow93}
  Downes, R. A., \& Shara, M. M., 1993, PASP, 105, 127
\bibitem[Hall (1989)]{hal89}
  Hall, D. S., 1989, SSRv, 50, 219
\bibitem[Kraft (1964)]{kra64}
  Kraft, R. P., 1964, ApJ, 139, 457
\bibitem[Lanza \& Rodon\`{o} (2002)]{lan02}
  Lanza, A. F., \& Rodon\`{o}, M., 2002, A\&A, 390, 167
\bibitem[Lanza (2005)]{lan05}
  Lanza, A. F., 2005, MNRAS, 364, 238
\bibitem[Lanza (2006)]{lan06}
  Lanza, A. F., 2006, MNRAS, 369, 1779
\bibitem[Liao \& Qian (2010)]{lia10}
  Liao, W. P., \& Qian, S. B., 2010, MNRAS, in press
\bibitem[Mumford \& Krzeminski (1969)]{mum69}
  Mumford, G. S. \& Krzeminski, W., 1969, ApJS, 18, 429
\bibitem[Mumford (1980)]{mum80}
  Mumford, G. S., 1980, AJ, 85, 748
\bibitem[North et al. (2000)]{nor00}
  North, R. C., Marsh, T. R., \& Moran, C. K. J., 2000, MNRAS, 313, 383
\bibitem[Pringle (1975)]{pri75}
  Pringle, J. E., 1975, MNRAS 170, 633
\bibitem[Robinson (1974)]{rob74}
  Robinson, E. L., 1974, ApJ, 193, 191
\bibitem[Shafter \& Wheeler (1986)]{sha86}
  Shafter, A. W., \& Wheeler, J. C., 1986, ApJ, 305, 261
\bibitem[Stover et al. (1981)]{sto81}
  Stover, R. J., Robinson, E. L., \& Nather, R. E., 1981, ApJ, 248, 696
\bibitem[Stumpff (1980)]{stu80}
  Stumpff, P., 1980, A\&AS 41, 1
\bibitem[Tout \& Hall (1991)]{tou91}
  Tout, C. A., \& Hall, D. S., 1991, MNRAS 253, 9
\bibitem[Verbunt \& Zwaan (1981)]{ver81}
  Verbunt, F. \& Zwaan, C., 1981, A\&A 100, 7
\bibitem[Warner (1995)]{war95}
  Warner B., 1995, Cambr. Astrophys. Ser. 28, Cataclysmic Variable Stars. Cambridge Univ. Press, Cambridge
\bibitem[Welsh et al. (2007)]{wel07}
  Welsh, W. F., Froning, C. S., \& Marsh, T. R., et al., 2007, PASP, 362, 241
\bibitem[Winter \& Sion (2003)]{win03}
  Winter, L., \& Sion, E. M., 2003, ApJ, 582, 352

\end{thebibliography}
\end{document}